\documentclass[preprint,authoryear,12pt]{elsarticle}

\usepackage{epsfig}
\usepackage{multirow}

\usepackage{amssymb}

\usepackage[ps2pdf,%
a4paper=true,%
breaklinks=true,%
colorlinks=true,%
pdfauthor={First Author et al.},%
pdftitle={Template for manuscripts in Advances in Space Research}%
]{hyperref}

\journal{Advances in Space Research}

\begin{document}

\begin{frontmatter}



\title{SENSING THE EARTH'S LOW IONOSPHERE DURING SOLAR FLARES \\ USING VLF SIGNALS AND GOES SOLAR X-RAY DATA}


\author[label1]{Aleksandra Kolarski\corref{cor}}
\address[label1]{Institute for Geophysics, Batajni{\v c}ki drum 8, 11000 Belgrade, Serbia}
\cortext[cor]{Corresponding author}
\ead{aleksandrakolarski@gmail.com}


\author[label2]{Davorka Grubor}
\address[label2]{Faculty of Mining and Geology, Physics Cathedra, University of Belgrade, Dju{\v s}ina 7, 11000 Belgrade, Serbia}
\ead{davorkag@eunet.rs}


\begin{abstract}
An analysis of D-region electron density height profile variations,
induced by four isolated solar X-ray flares during period from
September 2005 to December 2006, based on the amplitude and the phase delay
perturbation of 22.1 kHz signal trace from Skelton (54.72 N, 2.88 W)
to Belgrade (44.85 N, 20.38 E), coded GQD, was carried out. Solar
flare data were taken from NOAA GOES12 satellite one-minute
listings. For VLF data acquisition and recordings at the Institute
of Physics, Belgrade, Serbia, the AbsPAL system was used. Starting
from LWPCv21 code \citep{F98}, the variations of the
Earth-ionosphere waveguide characteristic parameters, sharpness and
reflection height, were estimated during the flare conditions. It
was found that solar flare events affected the VLF wave propagation in
the Earth-ionosphere waveguide by changing the lower ionosphere
electron density height profile, in a different way, for different
solar flare events.
\end{abstract}


\end{frontmatter}

\parindent=0.5 cm

\section{Introduction}

Production of electrons in D-region (which extends in altitude
between a range of heights from 50 km to 90 km) in unperturbed
ionospheric conditions is mainly related to the photoionization
processes, UV Lyman-$\alpha$ spectral line 121.6 nm, EUV spectral
lines ranging from 102.7 nm to 118.8 nm and galactic cosmic rays.
Different space phenomena, like X-ray solar flares \citep{TRC05}, solar
eclipse \citep{CPSMRBMKBC12}, CME \citep{bal08}, $\gamma$-ray
bursts \citep{ina07,cha10}, solar terminator \citep{nina13}, affect the changes
in intensity of the incidence of radiation in D-region and consequently change the
electron density time and space distributions.

During the flare events, the ionization in the lower ionosphere, induced by
electromagnetic radiation from the solar X-ray range (0.1-0.8 nm), significantly
exceeds the ionization of all the regular ionization factors
(such as Lyman-$\alpha$ spectral line 121.6 nm and cosmic rays) and causes
photoionization of neutral constituents in the lower ionosphere, becoming a major
source of ionization in this region \citep{WP65}.
Electron density increases as a result of
additional ionization of the lower ionosphere constituents and thus
changes the lower ionosphere electron density height profile,
affecting the Earth-ionosphere waveguide characteristics \citep{M74}.

Electron density in the ionosphere can be determined according to many different methods,
including the rocket probes, radar measurements and the technique using radio wave signals
\citep{BOPU06,MF07,ZGS07,CMSPBCBRMNYKKGGCPI12}. As in many other
manuscripts \citep{PC10,BC13,NCSZ12b,tho11}, for monitoring of the lower ionosphere,
Very Low Frequency (VLF) radio signals are used in this paper as well. One of the widely
used procedures for electron density calculations from VLF data is based on the application
of Wait theory and LWPC (Long Wave Propagation Capability \citep{F98})
numeric routine code \citep{MT00,GSZ08,ZGS07,tho11,KGS11,NCSS11,NCSS12a,NCSZ12b}.

Propagating VLF signal amplitude and phase delay, otherwise stable under undisturbed
solar conditions \citep{T93,MT00}, undergo perturbations since the
VLF signal propagation parameters change as a consequence of electron density
increase in the lower ionosphere induced by the solar X-ray flare events.
Electron production rate coefficient can be considered as directly proportional to
ionizing X-ray radiation intensity \citep{R72,B88} (a typical representation of
electron density responses to the incidence of X-ray radiation at different altitudes is
given in \citep{NCSZ12b}) and that is the reason why it is possible to draw
conclusions about the simultaneous changes of electron density height profiles
in D-region by analyzing VLF signals recorded
during the solar flare events, as it is shown in this paper.

The Absolute Phase and Amplitude Logger (AbsPAL) receiving system,
located at Institute of Physics in Belgrade, was used for receiving,
monitoring and for the storage of amplitude and phase delay of VLF data
on frequency 22.1 kHz (GQD signal emitted from Skelton, UK). The
phase delay and amplitude signal perturbations on GQD/22.1 kHz
signal traces, produced by C and M class isolated X-ray solar flare
events at equinox, winter and summer season, were studied and are
presented in this paper.

\section{Results and Discussion}

\begin{figure}
\begin{center}
\includegraphics*[width=8cm,angle=0]{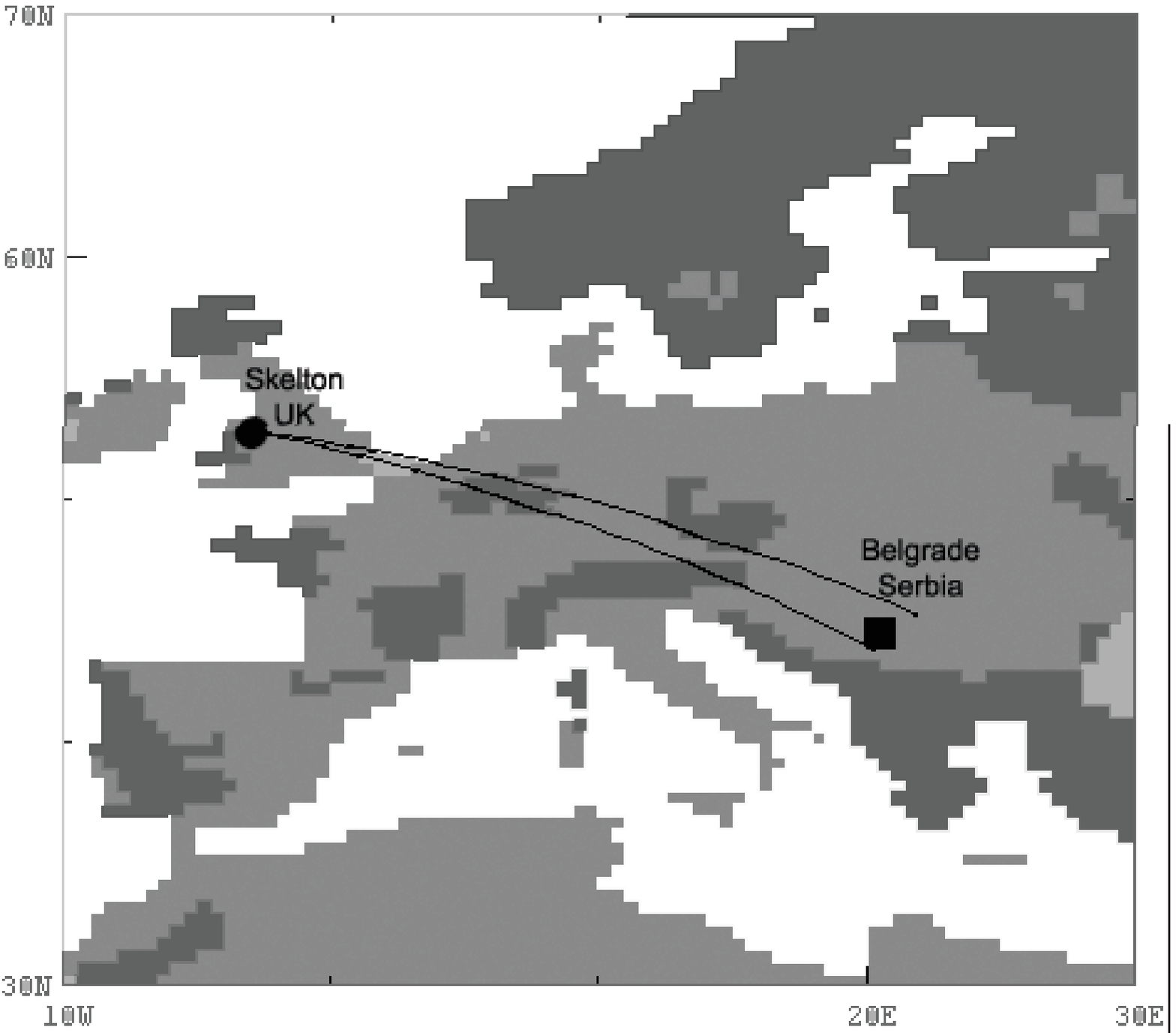}
\end{center}
\caption{Possible paths for GQD/22.1 kHz VLF signals emitted from
Skelton (UK) toward Belgrade (Serbia).} \label{figure1}
\end{figure}

The GQD signal propagates WNW-ESE along a short Great Circle Path
($D_{GCP}$ = 1980 km) and mostly an overland path (Figure
\ref{figure1}). The X-ray (0.1-0.8 nm) flare events, according to
NOAA GOES12 one-minute data listings, used in the following
exemplary analysis are given in Table \ref{table1}.

\begin{figure}
\begin{center}
\includegraphics*[width=10cm,angle=0]{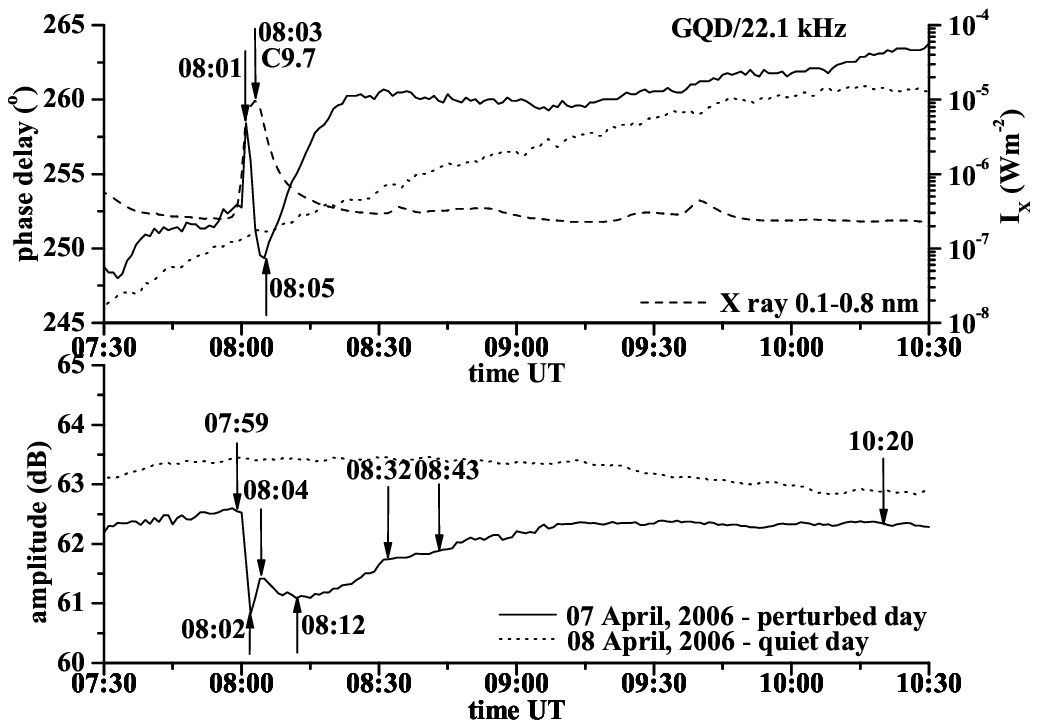}
\end{center}
\caption{GQD signal perturbation during C9.7 class X-ray solar flare
event.} \label{figure2}
\end{figure}

\begin{figure}
\begin{center}
\includegraphics*[width=10cm,angle=0]{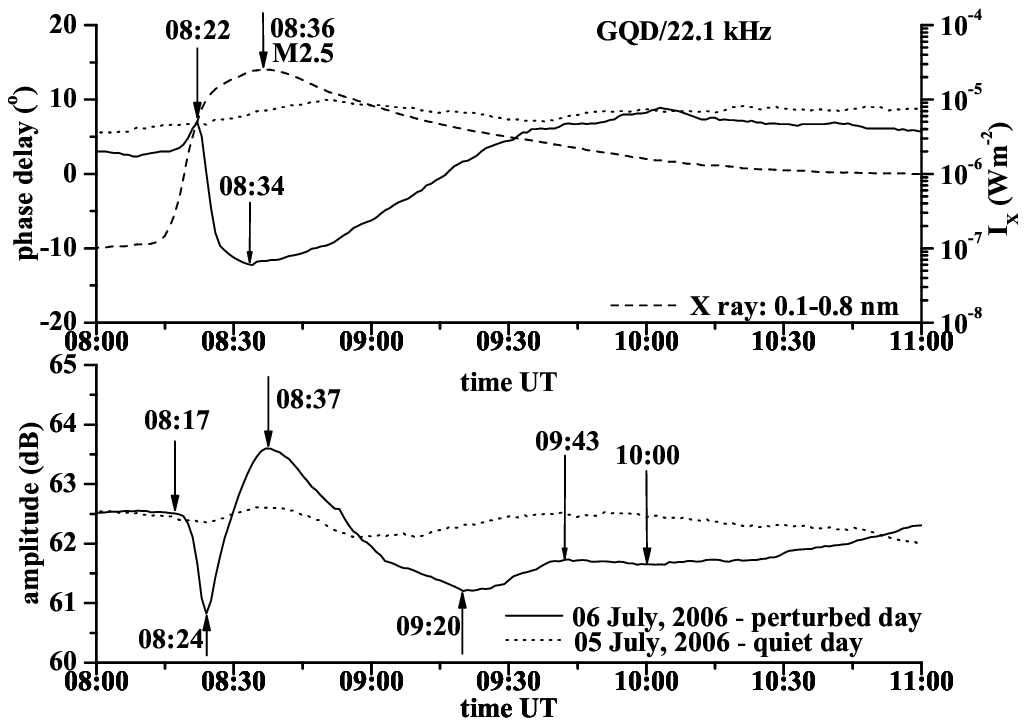}
\end{center}
\caption{GQD signal perturbation during M2.5 class X-ray solar flare
event.} \label{figure3}
\end{figure}

\begin{figure}
\begin{center}
\includegraphics*[width=10cm,angle=0]{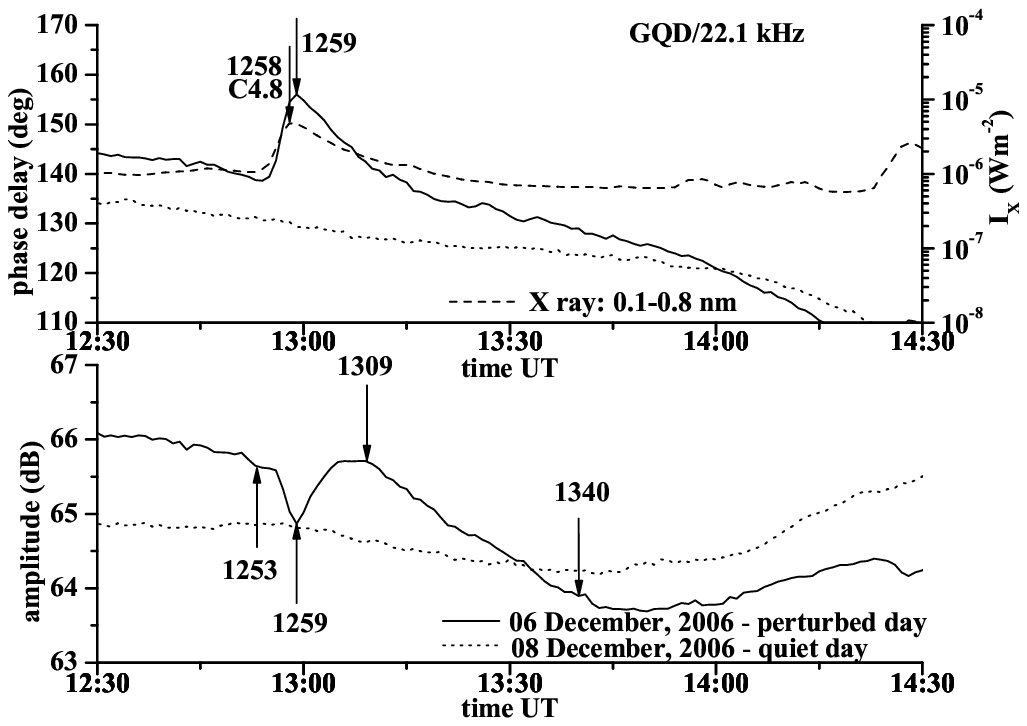}
\end{center}
\caption{GQD signal perturbation during C4.8 class X-ray solar flare
event.} \label{figure4}
\end{figure}

\begin{figure}
\begin{center}
\includegraphics*[width=10cm,angle=0]{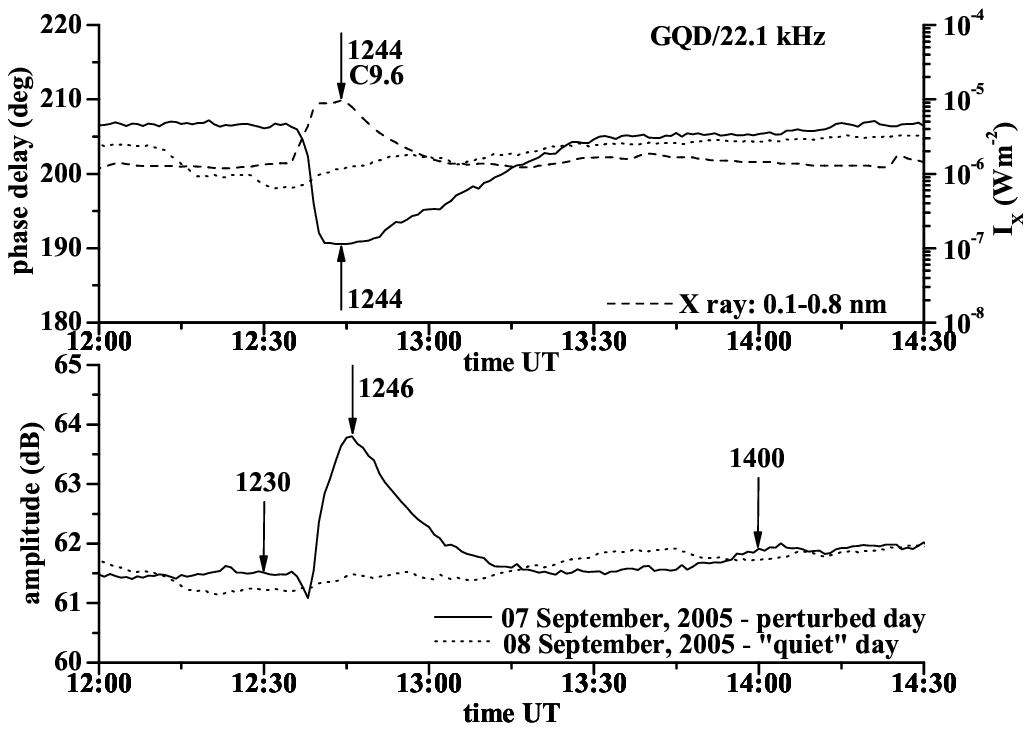}
\end{center}
\caption{GQD signal perturbation during C9.6 class X-ray solar flare
event.} \label{figure5}
\end{figure}

The VLF signal perturbations related to analyzed solar flare events,
observed on GQD/22.1 kHz signal traces, in periods enclosing the
flare events on the perturbed days, are shown in Figures
\ref{figure2}-\ref{figure5} using the solid lines (phase delay on upper
plots and amplitude on lower plots), respectively. The diurnal phase
delay and amplitude variations for GQD/22.1 kHz signal traces, in
the same daytime periods, but on the quiet days considered, are
added to the upper and lower panels of Figures
\ref{figure2}-\ref{figure5} and are presented with the dotted lines. The
X-ray irradiances during the periods of the flare events impact are
added to the upper panels of Figures \ref{figure2}-\ref{figure5}
(outer right axes) and are presented using the dashed lines.
Characteristic signal states during the analyzed flare events are marked
by arrows.

As evident from Figures \ref{figure2}-\ref{figure5}, solar X-ray
flares caused phase delay and amplitude perturbations on GQD signal
traces. However, the "pattern" of perturbations is not the same for
these events. The reason for such a behavior is that analyzed signals
are propagating through the different waveguides due to diurnal and
seasonal changes of the lower ionosphere and because of the different
events' characteristics. The common feature for all four events is
peak amplitude time delay, $\Delta$$t$ = 1 - 2 min, after the peak
of X-ray irradiance, and it was attributed to the "sluggishness" of
the ionosphere in reaching the peak electron density in D-region,
induced by flare and caused by recombination processes
\citep{M74,A53,ZGS07,NCSZ12b}. The amplitude and phase
delay GQD signal perturbations during analyzed flare events
have oscillatory character, and the type of oscillations is related
to the class of the observed solar flare event \citep{GSZ08}.

The incidence of the X-ray radiation in the Earth's ionosphere
during the solar flare causes not only an enhancement of the maximum
electron density, but it also changes the distribution of ionization
from an upper to a lower edge of the D-region. The propagation
model \citep{WS64} considered the electron density $N_e$ (m$^{-3})$
in the waveguide at the altitude $z$ (km), by two
parameters: reflecting edge sharpness, denoted by $\beta$ (km$^{-1}$) and
reflecting edge height, denoted by $H'$ (km):

\begin{equation}
N_e (z, H', \beta) =
1.43\cdot10^{13}e^{-0.15H'}e^{(\beta-0.15)(z-H')}~.
\end{equation}

This model has been used to simulate VLF propagation through the
Earth-ionosphere waveguide at regular conditions \citep{T93,MT00},
as well as for the conditions corresponding to the flare peak
irradiance \citep{MT04,TRC05}. The Earth-ionosphere waveguide was
modeled for several characteristic moments during each flare
event and the results obtained are in line with VLF signal
measurements.

By means of a LWPCv21 code, the propagation paths of VLF wave
on frequency 22.1 kHz were simulated with a goal to estimate the
best fitting pairs of parameters ($\beta$, $H'$) to yield values closest
to a real measured phase delay and the amplitude at the Belgrade
receiver site, for each characteristic state of each considered flare event.

LWPC is a set of several separate programs, each designed for
implementation of specific operations. LWPM program (Long Wave
Propagation Model) implies a standard model of the ionosphere with an
exponential conductivity increase with height. LWPC program takes
values $\beta$ = 0.30 km$^{-1}$ and $H'$ = 74 km, as standard values
for regular (unperturbed) daytime ionosphere conditions. In order to
simulate propagation conditions held in the perturbed waveguide during
the flare impact, it is necessary to modify a propagation model, using
LWPC subprogram REXP (Range Exponential Model) which calculates the
phase delay and the amplitude of the given VLF signal, depending on a chosen
GCP path and corresponding pairs of ($\beta$, $H'$)
parameters as impute parameters defined by the user. If pairs of ($\beta$, $H'$)
parameters are correctly defined, then the numerically simulated VLF
signal phase delay and amplitude values will be very close to the
measured values of VLF signal for a given signal trace.

Since parameters ($\beta$, $H'$) change along the signal trace,
in case of a GQD signal which propagates along the path 1980 km
long from a transmitter to a receiver, it corresponds to one
time zone. For this purpose, the constant "average value" of otherwise
variable parameters $\beta$, $H'$ was chosen and used along the
whole trace for the analyzed signals, depicting "average ionospheric
conditions" held along the whole trace for each simulation.

Perturbed phase delay and amplitude values, at characteristic states
during the flare impact, were denoted as $P_{flare}$ and $A_{flare}$,
depending on each particular case, while corresponding phase delay
and amplitude regular values at quiet days were denoted as $P_{reg}$
and $A_{reg}$. For getting perturbed values of simulated phase delay
and amplitude, $P_{flare}$ and $A_{flare}$, for all flare events,
perturbed values of ($\beta$, $H'$) pairs of parameters defined
using REXP routine, were used. For this purpose, the procedure of
succeeding probe iterations was used. Depending on type of
perturbation, phase delay and amplitude perturbations, $\Delta$$P$
($^{\rm{o}})$ and $\Delta$$A$ (dB), can be positive or negative
values, and were calculated as: $\Delta$$P$ = $P_{flare}$ -
$P_{reg}$, and $\Delta$$A$ = $A_{flare}$ - $A_{reg}$.

The calculated amplitude and phase delay values obtained by LWPCv21
code are in good agreement with the measured values at the Belgrade receiver
site. Therefore, it can be further assumed that numerically modeled
signals had been transmitted in modeled ionospheric conditions which
are in good agreement with real ionospheric conditions held at the
time and measured at the place of the receiver site. The measured phase
delay and amplitude perturbations and estimated corresponding
parameters ($\beta$, $H'$) and corresponding electron densities at
74 km, calculated using (1) at three characteristic moments that
correspond to the unperturbed (preflare) state, perturbed flare state
and the "recovered" postflare state during the analyzed flare events, are
given in Table \ref{table2}.

The errors introduced by the technique used have been critically examined
to place the uncertainty of the results arrived at between 10\% and 20\%.
Also, deviations in determining electron concentrations using different
models vary for different altitudes. According to the analysis for perturbed
flare state ($I_{xmax}$) given in \citep{GSZ08}, electron density ratios
are within one order of magnitude, while in case of unperturbed flare
state given in \citep{nina13submitted} electron density ratios are smaller.
There are numerous techniques, some more precise, for determining ionospheric
parameters and electron densities that include complicated ionospheric
chemistry and sophisticated numerical codes \citep{BC13,NCSZ12b,ZGS07,PBMPC13,T10},
but taking into account that errors related to determining of ionospheric
parameters by using different methods are about one order of magnitude (factor 10), for the
purpose of qualitative analysis conducted in this work, results obtained
by applied Wait theory and LWPC numeric code are quite satisfactory.

\begin{figure}
\begin{center}
\includegraphics*[width=10cm,angle=0]{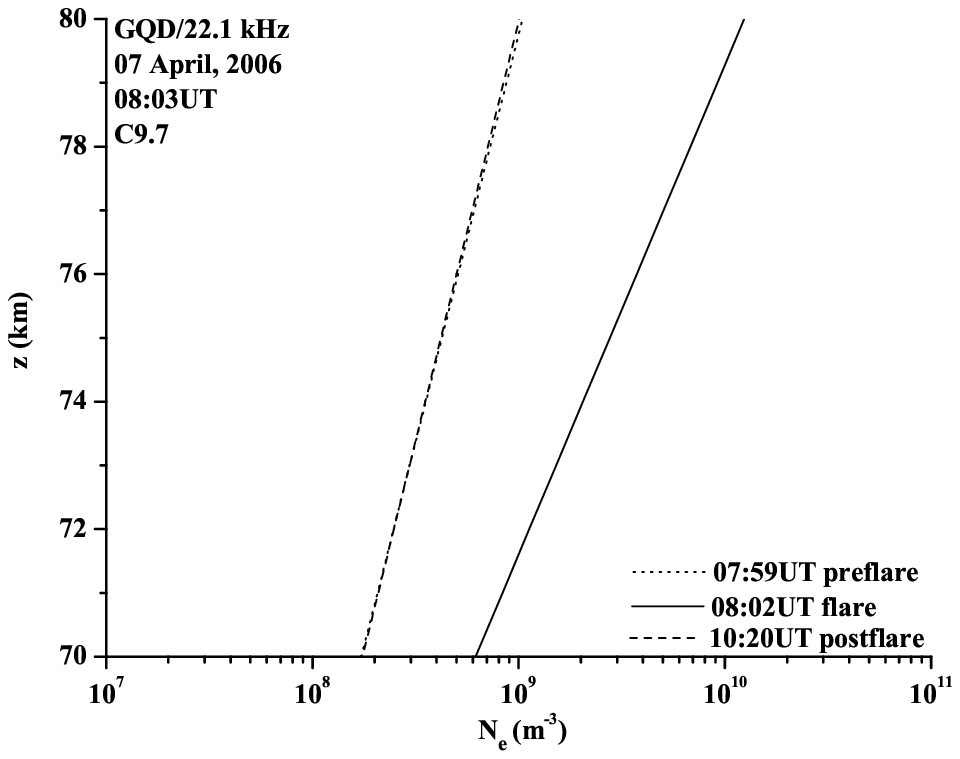}
\end{center}
\caption{Electron density height profiles during C9.7 class X-ray
solar flare event.} \label{fig:figure6}
\end{figure}

\begin{figure}
\begin{center}
\includegraphics*[width=10cm,angle=0]{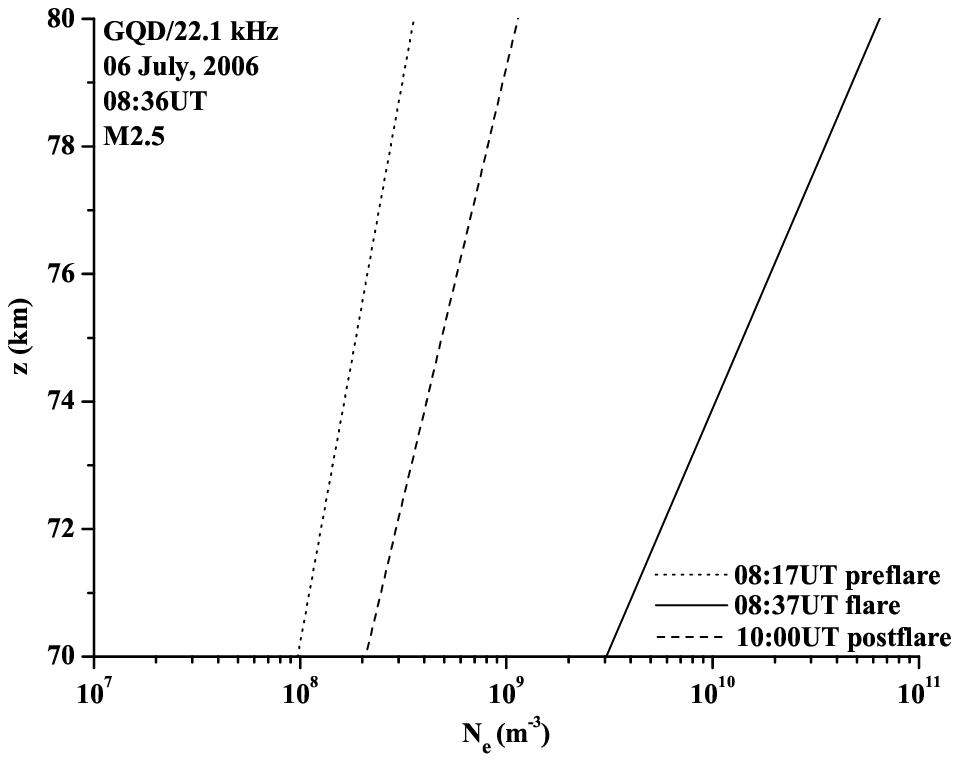}
\end{center}
\caption{Electron density height profiles during M2.5 class X-ray
solar flare event.} \label{fig:figure7}
\end{figure}

\begin{figure}
\begin{center}
\includegraphics*[width=10cm,angle=0]{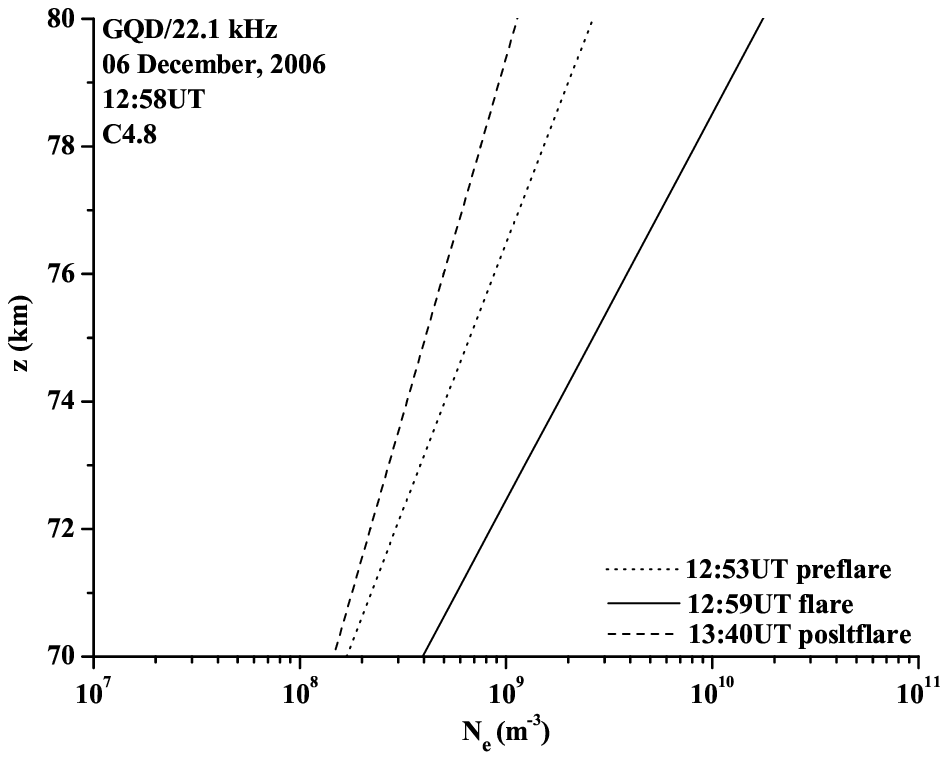}
\end{center}
\caption{Electron density height profiles during C4.8 class X-ray
solar flare event.} \label{fig:figure8}
\end{figure}

\begin{figure}
\begin{center}
\includegraphics*[width=10cm,angle=0]{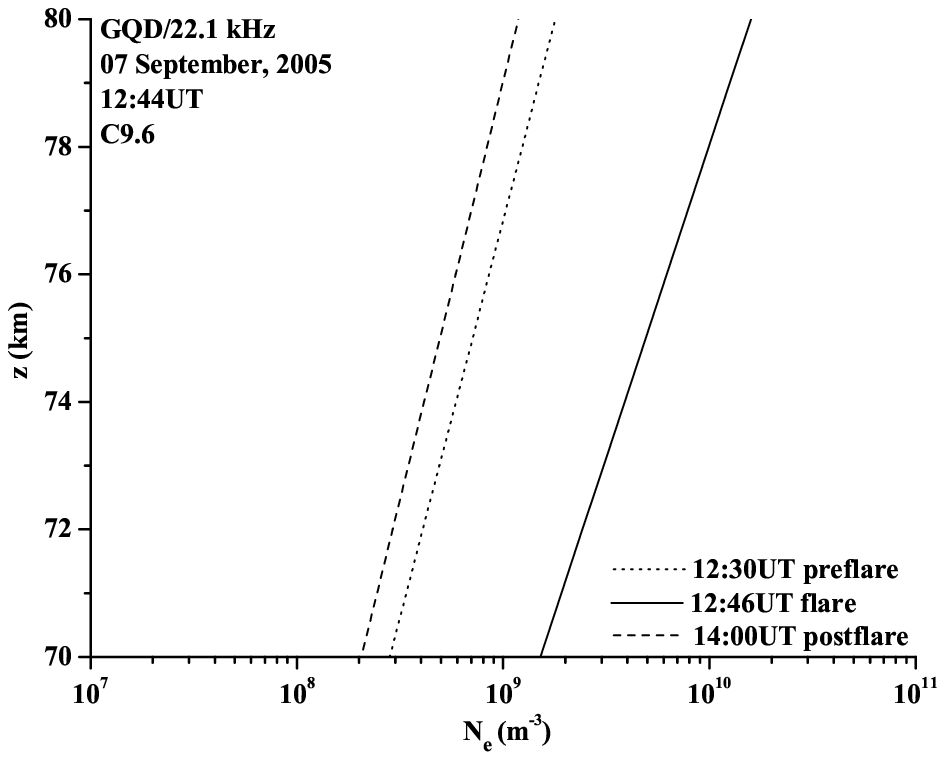}
\end{center}
\caption{Electron density height profiles during C9.6 class X-ray
solar flare event.} \label{fig:figure9}
\end{figure}

For all characteristic times (marked with arrows in Figures
\ref{figure2}-\ref{figure5}) during the analyzed flare events, the
vertical electron density height profiles through ionospheric
D-region (50 - 90 km altitude range) were calculated using (1).
Corresponding $N_e$ at altitude 74 km for all analyzed
characteristic times are given in Table \ref{table2}.
Since the GQD signal amplitude and phase delay perturbations
especially in case of C9.7 and M2.5 X-ray flare events (07 April,
2006 and 06 July, 2006, respectively), are quite complex, for a
clearer view only changes of the electron density height profiles
related to unperturbed preflare state (dotted lines), perturbed
flare state (solid lines) and "recovered" postflare state (dashed
lines) of ionospheric conditions for each flare event are shown in
Figures \ref{fig:figure6}-\ref{fig:figure9}. The changes of electron
density height profiles for GQD signal traces at ionospheric
D-region bottom (50 km) and at ionospheric D-region top (90 km)
should be taken with caution, because of the possible failure of the
model at the D-region boundaries. Nevertheless, the changes of
electron density height profiles at 74 km altitude are realistic,
and $N_e$(74 km) for the flare events considered is in line with
results of other studies \citep{ZGS07,MT04,KGS11,NCSS11,NCSS12a}.

Amplitude and phase delay VLF signal values computed
by means of LWPC code, which have been chosen as the
best fitting to the real measured amplitude
and phase delay at the receiver site, are the last values
in the series of values that the program code computes
for every point on the signal path, based on the corresponding
propagation model of the wave package inside the waveguide.
If the chosen pair of ($\beta$, $H'$) parameters reproduces the
values of a measured amplitude and phase delay at the receiver
site well, then all the computed values of amplitude and the phase delay
for every point on the signal path can be considered reliable.
In other words, for every moment of the change in the waveguide
state conditions, the signal amplitude and phase
delay changes depending on the location along the GCP,
can be simulated. One should bear in mind that the chosen
pair of ($\beta$, $H'$) parameters
depicts "average ionospheric conditions" in the waveguide
and in case of GQD signal trace, the GCP distance is
1980 km and it covers fewer than two time zones.

The GQD signal amplitude and phase delay changes along
the GCP, caused by the four X-ray solar flare events considered,
were analyzed for preflare, flare and postflare states. The changes
of the amplitude and phase delay at the site of the main modal
minimum and at the site of the receiver were analyzed and
also the morphology and the changes of the location of the
main modal minimum during considered solar flare events.
For GQD signal trace analyzed in this paper, the most prominent
changes in the waveguide state, caused by the incidence of X-ray
radiation during the solar flare events, occur at the location of
the main modal minimum and not at the receiver site. During all
the solar flare events considered (especially in case of C9.7,
M2.5 and C9.6 flare events), the main modal minimum of GQD
signal becomes mitigated and moves toward the transmitter,
although in case of the weak C4.8 flare event changes of the
main modal minimum characteristics and location are negligible.

Simulation results of GQD signal amplitude and phase delay
changes along the GCP are given in Table \ref{table3}, where
symbol $\uparrow$ denotes the main modal
minimum mitigation in the flare state, $r$ is the ratio between the
simulated GQD signal amplitude at the main modal minimum
in postflare and preflare state: $r = A_{simpostflare}/A_{simpreflare}$,
$D$ (km) is the distance between the main modal minimum and the transmitter in preflare
($D_{preflare}$), flare ($D_{flare}$) and postflare ($D_{postflare}$) state,
$\Delta$$D$ (km) is the difference in the main modal minimum location between
postflare and preflare state: $\Delta$$D$ = $D_{postflare}$ - $D_{preflare}$
and $\Delta$$D_{f}$ (km) is the difference in the main modal minimum location
between the flare and preflare state: $\Delta$$D_{f}$ = $D_{flare}$ - $D_{preflare}$.

In the flare state in case of C9.7 flare event, the amplitude and phase
delay minimum tended to form at the receiver site (denoted with "$\ast\ast$" in
Table \ref{table3}) which additionally lowered the registered amplitude
and phase delay. In the flare state in case of M2.5 flare event, at
1760 km away from the transmitter, an additional modal minimum
was formed, which led to a slight increase in the amplitude and a
decrease in the phase delay registered at the receiver site. In the
flare state in case of C9.6 flare event, at 1760 km away from the
transmitter, an additional modal minimum was formed, which led
to an increase in the amplitude followed by a decrease
in the phase delay registered at the receiver site. Flare event of
C4.8 class did not significantly change the waveguide
characteristics and it relatively weakly affected GQD signal
amplitude and the phase delay variation along the GCP.

After the impact of C9.7 and C9.6 solar flare events, the
propagation conditions in the GQD signal waveguide returned
to the regular and established signal amplitude and phase
delay variation along the GCP in postflare state was almost
identical as in the preflare state (parameters $r$ and $\Delta$$D$ in
Table \ref{table3}). In the state of recovered ionospheric
conditions after the C4.8 solar flare event impact, a
secondary modal minimum at 760 km away from the
transmitter became more pronounced than the modal
minimum considered as the main in preflare and post
flare states (denoted with "$\ast$" in Table \ref{table3}).
It should be noted that the recovery for this flare event lasted for
41 minutes and also that the ionosphere before this flare
event occurred had already been perturbed by a strong
flare event of M6 class which occurred earlier that day at
08:22UT, so that the ionosphere actually got back to its
regular state after the C4.8 solar flare event, at
12:59UT. Only M2.5 flare event, also the strongest
of the flare events considered, had left more
lasting effects on the GQD signal waveguide.

Deviations from the characteristic VLF signal amplitude and
phase delay variation scheme indicate that the disturbance
in the Earth-ionosphere waveguide took place. Such changes
of the waveguide characteristics can cause different types of
changes in the amplitude and phase delay, with more or less
complicated patterns and both the amplitude and phase delay
(the phase angle of signal reception) can increase or decrease.
If the disturbance occurs during daytime, its cause is most
likely the incidence of solar X-ray radiation in the lower
ionosphere. The incidence of solar X-ray radiation in the
upper VLF waveguide boundary can cause a) the "sharpening"
and the descending of the lower edge of the ionosphere
(the upper boundary of the VLF waveguide) due to electron
density increase, or b) increasing of the total electron content
in the whole D-region without the "sharpening" of the lower
edge of the ionosphere.

In case a) the so called "mirror" type of VLF signal reflection
occurs, because of the reflecting surface, in terms of
electrical conductivity, behaves as metal.
Energy dissipation of incidence VLF wave is negligible,
even less than that in unperturbed ionospheric conditions.
Such disturbances in the waveguide cause an increase of
registered VLF signal amplitude. The reflected VLF signal
amplitude changing follows the X-ray radiation intensity
changes during the flare event: the amplitude relatively
rapidly increases until it reaches its maximal value and
then gradually decreases until it reaches the value that
corresponds to the unperturbed value of the amplitude
before the flare event impact. In case b) VLF signal
penetrates into the D-region up to the altitude where
VLF signal frequency equalizes to plasma frequency
which is the site of VLF signal reflection. Along the part
of the path which is inside D-region,
refraction coefficient of VLF signal varies from one point
to another with constant energy dissipation, depending
on VLF signal frequency (deviant absorption, \citep{B88}).
Reflected VLF signal leaves lower ionospheric boundary
with energy lower than that of incident signal.
In this case, rapid electron density increase in the lower
ionosphere due to the incidence of X-ray radiation during solar
flare is followed by rapid VLF signal amplitude decrease until the
amplitude reaches its minimal value and then VLF propagation
parameters go back to the values that correspond to
unperturbed regular conditions in the lower ionosphere.

As a result of the phase trajectory shortening over the GCP due to
VLF signal reflection height lowering during the flare event impact,
the location of the main modal minimum (and of all others, as well) on
the signal trace is moved toward the transmitter site. There are
fewer wavelengths on the shortened trajectory,
which correspond to a smaller delay in phase, therefore a higher
phase angle. On the other hand, VLF signal penetrations into the
D-region cause a phase trajectory elongation by each VLF signal
reflection from the ionosphere, which correspond to a greater
delay in the phase, therefore a smaller phase angle.

The value of the registered VLF signal amplitude is also affected by the
interference modes layout along the GCP and thus by vicinity of the
modal minima and maxima to the receiver site as well. The increase
in the amplitude, which is characteristic for the case a), is going to be
more pronounced if the receiver is placed in the
vicinity of a modal maximum than if it is placed in the vicinity
of a modal minimum. Moreover, if the receiver is placed in the
vicinity of a modal maximum, the decrease in amplitude, which
is characteristic for the case b), is going to be less pronounced than
if the receiver is placed in the vicinity of a modal minimum.
The sign of the phase delay change (phase trajectory shortening
or elongating) depends on the interference modes layout along
the GCP, too, but also depends on the length of the VLF signal path.

The amount of amplitude and phase delay changes during flare state
depends on the redistribution of the local minima and maxima
locations along the signal trace compared to the regular unperturbed
state in the waveguide. The distance between the transmitter and the
receiver along the GCP (the number of signal reflections in the waveguide)
and the incidence angle at the lower D-region boundary affect the behavior
of the signal, especially in the perturbed ionospheric conditions.
The phase trajectory modification is the result of both effects. In most
cases, GQD signal penetrates into the area of an increased electron
density above the lower D-region boundary, where the deviant energy
absorption takes place, which manifests at the receiver site as the GQD
signal amplitude decreases, compared to the preflare
and postflare states, as in case of C9.7 and C4.8 flare events.
GQD signal reflection of the "mirror" type occurs only in the case of
higher class X-ray solar flare events and it is only related to the
moments of maximal X-ray irradiance, such as in cases of
M2.5 and C9.6 flare events.

Although the effects on VLF signal amplitude and phase delay which
are induced by X-ray solar flare events can always be clearly noticed
and recognized, the flare events with the same characteristics (class,
duration, zenith  angle of incidence radiation) can cause different
types of amplitude and phase delay perturbations depending on
which VLF signal trace it is monitored. For example, all four flare
events which have been analyzed in this paper on GQD
signal, in case of NAA/24.0 kHz signal (the GCP distance 6540 km) are
characterized with "mirror" type of VLF signal reflection at the Belgrade
receiver site \citep{KGS11}. "Mirror" type of signal reflection is
characteristic for NAA signal received at the Belgrade receiver site
regardless of the flare event strength \citep{KGS11}.

\section{Conclusions}

The aim of this paper was an extended comparative qualitative
analysis regarding the behavior of D-region electron population
during the four analyzed X-ray solar flare events that impacted
the lower ionosphere VLF propagation on a GQD signal trace.
We analyzed patterns that occur on a GQD signal
during these solar X-ray flares depending on the X-ray
solar flare event's strength and the changes in amplitude and
phase delay modal minima and maxima movements along
the GCP caused by these X-ray solar flare events. We have
also analyzed the electron density change responses during the
whole time period that X-ray solar flares impacted transmissions
on GQD signal and electron density height profiles throughout
D-region, in the altitude domain around the signal reflection heights
where Wait theory can be applied in a reliable manner. The
application of Wait theory to the boundaries of ionospheric region
considered is somewhat unrealistic, thus the results concerning ionospheric
D-region bottom and top should be taken with caution and
were analyzed only in the sense to provide general insight into
electron population behavior, i.e. trend of changes. This is
significantly more important during the X-ray solar flare event's impact.

Each analyzed flare event impact on GQD signal amplitude and phase
delay is specific. In general, the amplitude decrease, due to deviant
energy absorption, followed by the phase delay increase, occurs as the
result of VLF signal reflection height lowering. Such perturbations
can be seen in cases of solar flare events, which occurred on 07 April,
2006 and on 06 December, 2006. Different type of perturbations took
place at flare peaks occurred on 06 July, 2006 and on 07
September, 2005. In these cases, after short-lasting decrease, the
amplitude increases until it reaches its maximal value above the regular
ionospheric conditions level, while simultaneously phase delay
decreases until it reaches its minimal value. It is clear that amplitude
increases due to the "mirror" type of VLF signal reflection, as
simultaneously reflecting height lowers and phase trajectory
shortens. But, instead of the decrease, increase of the delay in the
phase occured, so the GQD signal phase descends.
The amount of these changes depends on the solar flare event
class, and also on the modal extremes distribution along the GCP, therefore,
on electron density height profile at a certain moment and at a certain
location in the waveguide.

\section{Acknowledgments}
The authors wish to acknowledge insightful comments and suggestions and
valuable discussions on this topic with A. Nina and V. {\v C}ade{\v z}
in preparing this manuscript. The authors appreciate comments expressed
by all referees, which have led to significant improvements in the
contents of this paper.
The thanks are due to Ministry of Science and Technological
Development of Serbia, project III 44002.


\clearpage

\newcommand{\minitab}[2][l]{\begin{tabular}{#1}#2\end{tabular}}
\begin{table}
\begin{center}
\caption{The flare events below are observed on the analyzed GQD/22.1 kHz signal traces}
\label{table1}
\begin{tabular}{|c|c|c|c|c|}
\hline {\multirow{2}{*}{date}} & \multirow{2}*{\minitab[c]{time \\(UT)}}& {\multirow{2}{*}{class}}&  \multirow{2}*{\minitab[c]{$I_{xmax}$ \\ $(\rm{Wm}^{-2})$ }}& {\multirow{2}{*}{quiet day}}\\
& & & & \\
\hline 07 Apr. 2006 &  08:03 &  C9.7 &  $9.74\cdot10^{-6}$ &  08 Apr. 2006\\
\hline 06 July 2006 &  08:36 &  M2.5 &  $2.51\cdot10^{-5}$ &  05 July 2006\\
\hline 06 Dec. 2006 &  12:58 &  C4.8 &  $4.82\cdot10^{-6}$ &  08 Dec. 2006\\
\hline 07 Sep. 2005 &  12:24 &  C9.6  & $9.62\cdot10^{-6}$ &  08 Sep. 2005\\
\hline
\end{tabular}
\end{center}
\end{table}

\begin{table}
\begin{center}
\caption{Parameters characterizing GQD/22.1 kHz signal propagation
conditions for the X-ray solar flare events considered} \label{table2}
\begin{tabular}{|c|c|c|c|c|c|}
\hline \multirow{2}*{\minitab[c]{ flare\\ event }} & \multirow{2}*{\minitab[c]{ time \\(UT) }}& {\multirow{2}{*}{state}}&  \multirow{2}*{\minitab[c]{$\Delta$$A$ \\ (dB)}} &  \multirow{2}*{\minitab[c]{ $\Delta$$P$\\($^{\rm{o}}$) }} &  \multirow{2}*{\minitab[c]{$N_e$(74 km) \\ (m$^{-3}$)}}\\
& & & & & \\
\hline
\multirow{9}*{\minitab[c]{07 Apr. 2006 \\ 08:03UT \\ C9.7}} & 07:59 & preflare & -0.89 & 2.52 & 3.54$\cdot10^{8}$\\\cline{2-6}
& 08:01 & $P_{max}$ & -1.72 & 7.68 & 8.61$\cdot10^{8}$\\\cline{2-6}
& 08:02 & flare $A_{min1}$ & -2.62 & 4.93 & 2.05$\cdot10^{9}$\\\cline{2-6}
 & 08:04 & $A_{max1}$ & -1.98 & -1.69 & 5.90$\cdot10^{8}$\\\cline{2-6}
& 08:05 & $P_{min}$ & -1.99 & -1.78 & 5.90$\cdot10^{8}$\\\cline{2-6}
& 08:12 & $A_{min2}$ & -2.34 & 3.09 & 1.04$\cdot10^{9}$\\\cline{2-6}
 & 08:32 & $A_{max2}$ & -1.71 & 6.28 & 8.61$\cdot10^{8}$\\\cline{2-6}
& 08:43 & $A_{min3}$ & -1.55 & 4.48 & 6.56$\cdot10^{8}$\\\cline{2-6}
& 10:20 & postflare & -0.53 & 2.31 & 3.52$\cdot10^{8}$\\\hline
\multirow{8}*{\minitab[c]{ 06 July 2006\\ 08:36UT \\ M2.5 }} & 08:17 & preflare & 0.04 & -3.44 & 1.63$\cdot10^{8}$\\\cline{2-6}
& 08:22 & $P_{max}$ & -0.72 & 0.11 & 4.10$\cdot10^{8}$\\\cline{2-6}
& 08:24 & $A_{min1}$ & -1.54 & -6.46 & 1.56$\cdot10^{9}$\\\cline{2-6}
 & 08:34 & $P_{min}$ & 0.76 & -20.34 & 1.08$\cdot10^{10}$\\\cline{2-6}
& 08:37 & flare $A_{max1}$ & 1.0 & -20.25 & 1.03$\cdot10^{10}$\\\cline{2-6}
 & 09:20 & $A_{min2}$ & -1.11 & -6.54 & 8.25$\cdot10^{8}$\\\cline{2-6}
& 09:43 & $A_{max2}$ & -0.79 & -0.99 & 4.10$\cdot10^{8}$\\\cline{2-6}
& 10:00 & postflare & -0.79 & -0.44 & 4.10$\cdot10^{8}$\\\cline{2-6}
\hline \multirow{4}*{\minitab[c]{06 Dec. 2006 \\ 12:58UT \\ C4.8 }} & 12:53 & preflare & 0.8 & 7.44 & 5.06$\cdot10^{8}$\\\cline{2-6}
& 12:59 & flare $A_{min}$,$P_{max}$  & 0.06 & 26.71 & 1.80$\cdot10^{9}$\\\cline{2-6}
 & 13:09 & $A_{max}$ & 1.09 & 15.41 & 1.04$\cdot10^{9}$\\\cline{2-6}
& 13:40 & postflare & -0.33 & 5.51 & 3.31$\cdot10^{8}$\\\cline{2-6}
\hline \multirow{3}*{\minitab[c]{07 Sep. 2005\\ 12:44UT \\ C9.6 }} & 12:30 & preflare & 0.29 & 2.09 & 5.90$\cdot10^{8}$\\\cline{2-6}
& 12:46 & flare $A_{max}$,$P_{min}$ & 2.32 & -13.33 & 3.88$\cdot10^{9}$\\\cline{2-6}
& 14:00 & postflare & 0.19 & 1.36 & 4.14$\cdot10^{8}$\\\cline{2-6}
\cline{2-6} \hline
\end{tabular}
\end{center}
\end{table}

\begin{table}
\begin{center}
\caption{Simulation results of GQD signal propagation along the GCP for the X-ray solar flare events considered}
\label{table3}
\begin{tabular}{|c|c|c|c|c|}
\hline flare event class & C9.7 & M2.5 & C9.6 & C4.8\\
\hline $D_{preflare}$ (km) &  740 &  760 &  700 &  960\\
\hline $D_{flare}$ (km) &  700 &  600 &  600 &  940\\
\hline $D_{postflare}$ (km) &  740 &  720 &  720 &  960\\
\hline main mod. min. in flare state & $\uparrow$ & $\uparrow$ & $\uparrow$ & no change\\
\hline $r$ &  $\cong1$ &  $<1$  & $\cong1$ &  $\cong1$$\ast$\\
\hline $\Delta$$D$ (km) &  $\cong0$ &  $<0$  & $>0$ &  $\cong0$$\ast$\\
\hline $\Delta$$D_{f}$ (km) &  40 &  160  & 100 &  20$\ast$\\
\hline additional mod. min. &  no$\ast$$\ast$ &  yes  & yes &  no\\
\hline preflare ionospheric conditions &  regular &  regular & perturbed &  perturbed\\
\hline
\end{tabular}
\end{center}
\end{table}

\end{document}